\documentclass[aps,prl,showpacs,superscriptaddress,twocolumn,notitlepage,10pt]{revtex4-2}
\usepackage{amsmath}
\usepackage{amsfonts}
\usepackage{graphicx}
\usepackage{multirow}
\usepackage{braket}
\usepackage{chngcntr}
\usepackage{bm}

\usepackage{ulem}
\usepackage[dvipsnames]{xcolor}

\usepackage{natbib}
\usepackage[final=true]{hyperref}
\hypersetup{
   colorlinks=true,
    linkcolor=Maroon,
   filecolor=magenta,      
    urlcolor=cyan,
    citecolor=PineGreen,    
}

\newcommand{\ve}{\varepsilon}

\renewcommand{\d}{\dagger}
\newcommand{\be}{\begin{eqnarray}}
\newcommand{\ee}{\end{eqnarray}}
\newcommand{\nn}{\nonumber}

\newcommand{\om}{\omega}
\newcommand{\ketbra}[2]{\ket{#1}\!\bra{#2}}

\newcommand{\pcsadd}{Center for Theoretical Physics of Complex Systems, Institute for Basic Science (IBS), Daejeon, Korea, 34126}
\newcommand{\ustadd}{Basic Science Program, Korea University of Science and Technology (UST), Daejeon 34113, Republic of Korea}

\begin{document}

\title{
Mobility edges and fractal states in quasiperiodic Gross-Pitaevskii chains 
}

\author{Oleg I. Utesov}
\email[Electronic address: ]{utiosov@gmail.com}
\affiliation{\pcsadd}

\author{Yeongjun Kim}
\email[Electronic address: ]{yeongjun.kim.04@gmail.com }
\affiliation{\pcsadd}
\affiliation{\ustadd}

\author{Sergej Flach}
\email[Electronic address: ]{sergejflach@googlemail.com }
\affiliation{\pcsadd}
\affiliation{\ustadd}

\date{\today}

\begin{abstract}

We explore properties of a Gross-Pitaevskii chain subject to an incommensurate periodic potential, i.e., a nonlinear Aubry-Andre model. We show that the condensate crucially impacts the properties of the elementary excitations. In contrast to the conventional linear Aubry-Andre model, the boundary between localized and extended states (mobility edge) exhibits nontrivial branching. For instance, in the high-density regime, tongues of extended phases at intermediate energies penetrate the domain of localized states. In the low-density case, the situation is opposite, and tongues of localized phases emerge. Moreover, intermediate critical (fractal) states are observed. The low-energy phonon part of the spectrum is robust against the incommensurate potential. Our study shows that accounting for interactions, already at the classical level, lead to highly nontrivial behavior of the elementary excitation spectrum.

\end{abstract}

\maketitle

\textit{Introduction}---While many years passed since Anderson introduced the concept of localization~\cite{anderson1958}, it continues to attract significant attention for good reasons~\cite{anderson201050, evers2008anderson,beltukov2013,slevin2014, tikhonov2021anderson,conyuh2021,kati2021,vanoni2024}. 
The localization phenomenon is probed with new tools such as level spacing statistics (r-ratios)~\cite{oganesyan2007,atas2013distribution}, on new structures, including random regular~\cite{tikhonov2021anderson,vanoni2024} and irregular graphs~\cite{puschmann2015multifractal,bhattacharjee2024anderson}, models with correlated disorder~\cite{titov2005nonuniversality}, etc. 
Furthermore, a whole new field of many-body localization emerged~\cite{gornyi2005,basko2006,aleiner2010finite,michal2016finite,alet2018many}.
Nowadays, optical lattices of bosonic and fermionic atoms can be created, enabling the study and testing of important theoretical concepts, including localization. 
Experiments with disordered or quasiperiodic potentials have provided direct observation of Anderson localization~\cite{billy2008direct, roati2008anderson}, as well as the exploration of many-body localization through controlled interactions and disorder~\cite{schreiber2015observation}.

The Aubry-André (AA) model is one of the few models for which localization properties are known exactly~\cite{harper1955single,azbel1979,aubry1980analyticity,dominguez2019aubry}. This model consists of a simple nearest-neighbor hopping and an incommensurate on-site potential:
\be  \label{AA1}
  \mathcal{H}_{AA} &=& \sum_l \left( \ve_l a^\d_l a_l - J a^\d_{l+1} a_l - J a^\d_{l} a_{l+1} \right), \\ \label{AACorr}
  \ve_l &=& W \cos{\left( 2 \pi \beta l + \varphi \right)},
\ee
where the lattice parameter is set to unity, $\beta$ is an irrational number (otherwise, the Bloch theorem applies, and the system develops a band structure), and $\varphi$ is an arbitrary phase, which becomes unimportant in the thermodynamic limit $N \to \infty$.
Introducing the Fourier transform of the operators, one can see that the problem in reciprocal space is dual to the original. The self-dual point $W_\textrm{C} = 2 J$ separates the regimes of localized ($W > W_\textrm{C}$) and extended ($W < W_\textrm{C}$) states. At the critical point, all the states are fractal.
Note that $4 J$ is the bandwidth at $W = 0$.
The model has been experimentally realized in ultracold atom systems and photonic lattices, which allow for observing localization transitions and fractal wave functions~\cite{roati2008anderson}. Noteworthy, the AA model is also suitable for many-body generalizations~\cite{iyer2013,schreiber2015observation,mastropietro2015}.

In the present study, we consider an extension of the standard AA model~\eqref{AA1} by including interaction on the classical level, thereby making the connection to the physics of Gross-Pitaevskii lattices (see, e.g., Refs.~\cite{rasmussen2000,orzel2001squeezed,greiner2002quantum,polkovnikov2002,mithun2018}). 
Moreover, the considered model can be viewed as a toy one for studying many-body localization physics in the presence of an incommensurate periodic perturbation. 

We observe peculiar behavior of the Bogoliubov-de Gennes (BdG) excitations, which show non-trivial branching mobility edges, separating the extended, fractal, and localized modes. In a certain domain of the $W$-excitation energy plane, one can observe tongues of extended phase separated by localized states regions and ``reentrant'' behavior upon $W$ growth. We show that low-energy physics can be formulated in terms of acoustic modes subject to a renormalized AA potential governed by their wave number and $W$. This approach predicts that the low-energy modes remain delocalized regardless of the AA potential strength. This behavior contrasts with the standard AA model, where all eigenstates are localized when $W > W_\textrm{C}$. In the present case, one can introduce a quantity $W_\textrm{NL}$ which indicates when some modes, usually the highest-energy excitations, start to localize. In both low- and high-density regimes, we provide estimations for $W_\textrm{NL}$ which are numerically shown to work on a semi-quantitative level of accuracy.


{\it General formulae}---We consider the following Hamiltonian:
\be \label{ham1}
  \mathcal{H} = \sum_l \left[ \ve_l |\psi_l|^2 - J \left( \psi^*_{l+1} \psi_l + \psi^*_l \psi_{l+1} \right) + \frac{g}{2}|\psi_l|^4 \right], \nn \\
\ee
where $l=1...N$, $\psi_l$ is the condensate wave-function on a site $l$ (classical field), and $\ve_l$ is the Aubry-André on-site potential~\eqref{AACorr}. In addition to the kinetic energy governed by $J$, there is the nonlinearity proportional to $g$. Model~\eqref{ham1} with uncorrelated on-site disorder $\ve_l$
was considered in Ref.~\cite{kati2021}; we adapt some ideas of that study for our purposes. The equations of motion read
\be \label{eq_main}
  i \dot{\psi}_l = \frac{\partial \mathcal{H}}{\partial \psi^*_l} = \ve_l  \psi_l -J (\psi_{l+1} + \psi_{l-1}) + g |\psi_l|^2 \psi_l.
\ee
Notably, the Hamiltonian preserves the number of particles $A = \sum_l |\psi_l|^2$. Therefore, we introduce a crucial system parameter, the density of particles $a = A/N$. We distinguish two regimes of low and high densities $g a  \ll J$ and $ga \gg J$, respectively. It is pertinent to note that a variation of $a$ is equivalent to a rescaling of the interaction $g$.
In all our numerical calculations, we use dimensionless parameters and $J=g=1$~\cite{kati2021}. Unless stated otherwise, we set $\beta = (\sqrt{5}-1)/2$ (inverse golden ratio) and use its rational approximations in numerics.

\begin{figure}[t]
    \centering
    \includegraphics[width=\linewidth]{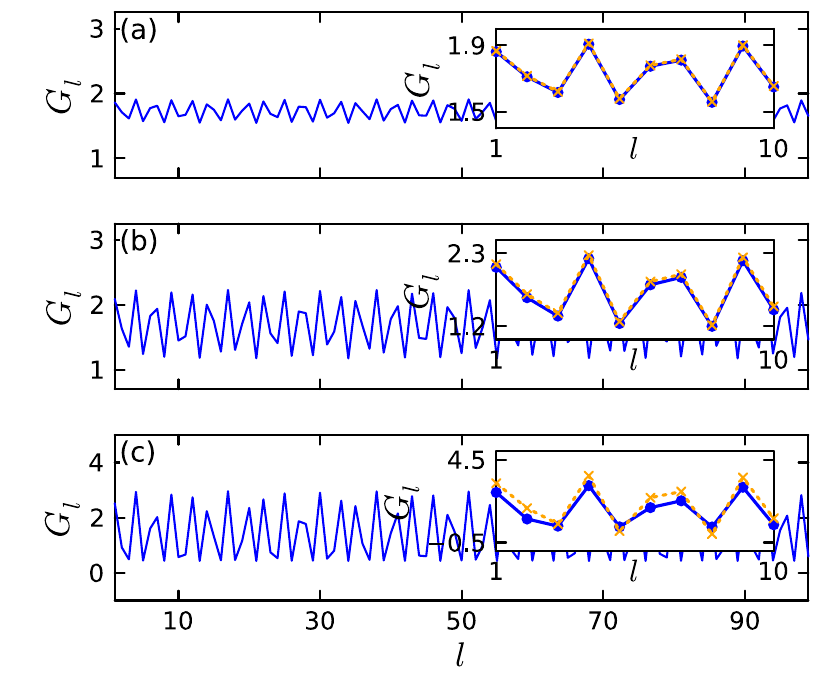}
    \caption{
    The ground state amplitudes (we use blue lines instead of discrete dots for presentation purposes) for the chain size $N=2584$. Here $a=3$ with (a) $W=1$,  (b) $W=3$, and (c) $W=10$. In the insets, we compare $G_l$'s and the linear response estimations~\eqref{gs} (dashed orange lines). Upon $W$ growth, deviations from the linear response become apparent.}
    \label{fig:gs}
\end{figure}

A standard trick of introducing the chemical potential $\mu$ and $\mathcal{H}^\prime = \mathcal{H} - \mu A$~\cite{kati2021} allows to formulate equations for the time-independent ground state $G_l$:
\be \label{eq_gs}
   (\ve_l -\mu) G_l -J (G_{l+1} + G_{l-1}) + g |G_l|^2 G_l =0,
\ee
which implicitly define $a$ as function of $\mu$ and other parameters (evidently, for $W = 0$ one has $ga = \mu + 2 J$). Next, we define the ground-state correlation field~\cite{kati2021} 
\be \label{eq_corr}
  \zeta_l = \frac{G_{l+1} + G_{l-1}}{G_l},
\ee
which is equal to $2$ in the absence of the AA potential. Using this field, we can rewrite equations for $G_l$'s as
\be 
  g G^2_l = J \zeta_l + \mu - \ve_l.
\ee 

For small $W \ll \textrm{max}(ga, J)$ the ground state varies following the AA potential
\be \label{gs}
  G_l \approx \sqrt{a} \left[ 1 - \frac{\ve_l}{2 g a +2 J( 1 - \cos{2 \pi \beta})}\right].
\ee
However, for $W \gtrsim \textrm{max}(ga, J)$, a strong nonlinear response is expected.

We compute the ground state using the gradient descent method (see Supplementary Material) with a constraint $\sum_l |\psi_l|^2 = A$ (the Lagrange multiplier corresponds to the chemical potential $\mu$).
Fig.~\ref{fig:gs} shows the numerically obtained $G_l$ in the high-density regime ($a = 3$) for different potential values $W = 1, 10, 40$ in panels (a)-(c), respectively.
The insets display a zoom-in for $1 \leq l \leq 10$, compared with the linear response prediction~\eqref{gs}. Evidently, the latter fails to accurately describe the ground state for strong AA potentials, namely for $W \gtrsim g a$ in the high-density regime. In the low-density regime, the same rule applies, but $W$ should be compared with $J$ (see End Matter).

\begin{figure}[t]
    \centering
    \includegraphics[width=1\linewidth]{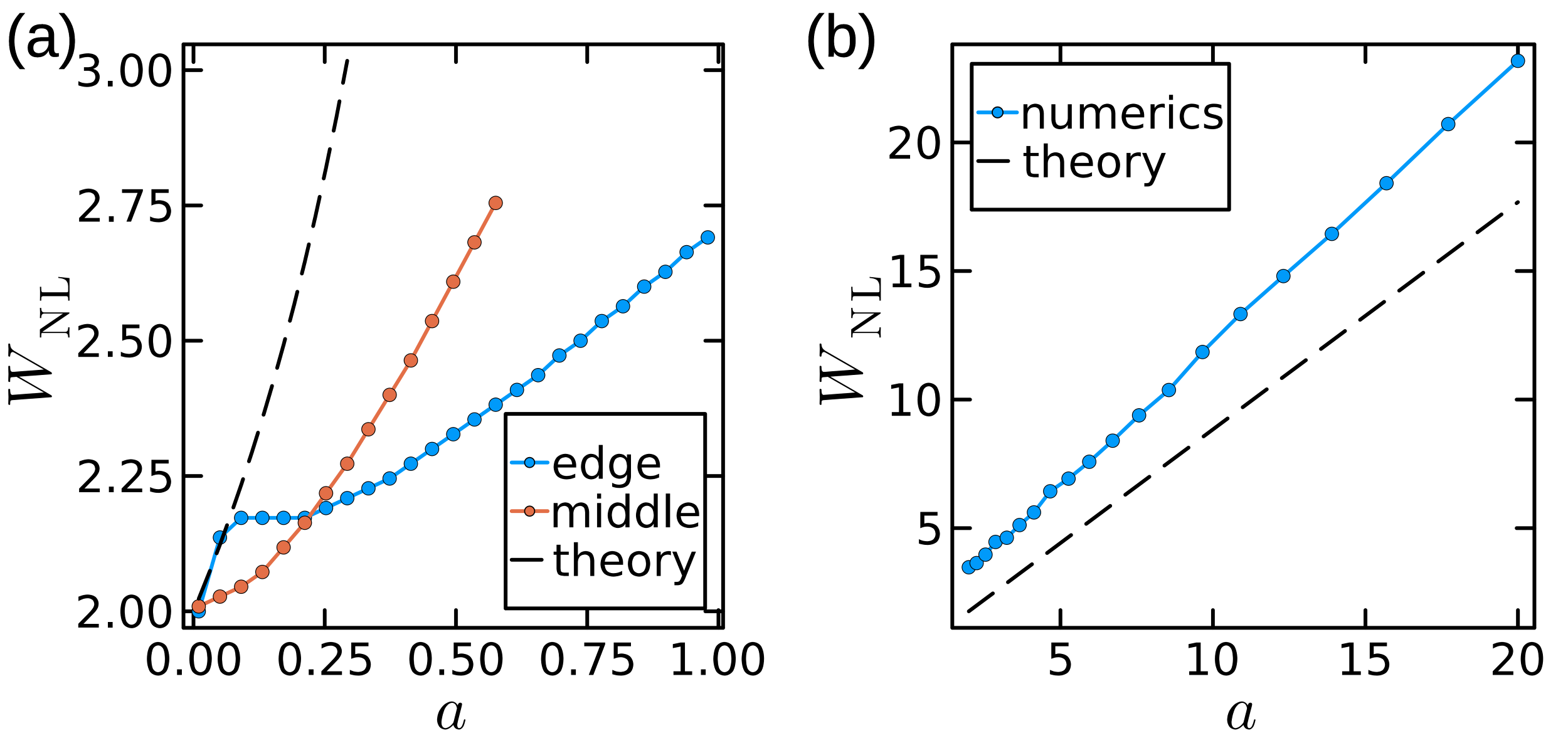}
    \caption{
    In the nonlinear Aubry-André model~\eqref{ham1}, $W_\textrm{NL}$ serves as an indicator when localized modes emerge upon $W$ growth. It depends on the particle density $a$. Here we plot it for low- (a) and high- (b) density regimes. 
    In (a), blue and red curves correspond to the emergence of the localized modes at the edge (\(\lambda \approx 5\)) and the middle (\(\lambda \approx 2.5\)) of the spectrum, respectively.
    In (b), only \(W_{\textrm{NL}}\) for the edge is presented. The dashed lines are theoretical predictions \eqref{LowDEst} and \eqref{HighDEst}.
    }
    \label{fig:wnl}
\end{figure}

\begin{figure*}[t]
    \centering
     \includegraphics[width=0.97\linewidth]{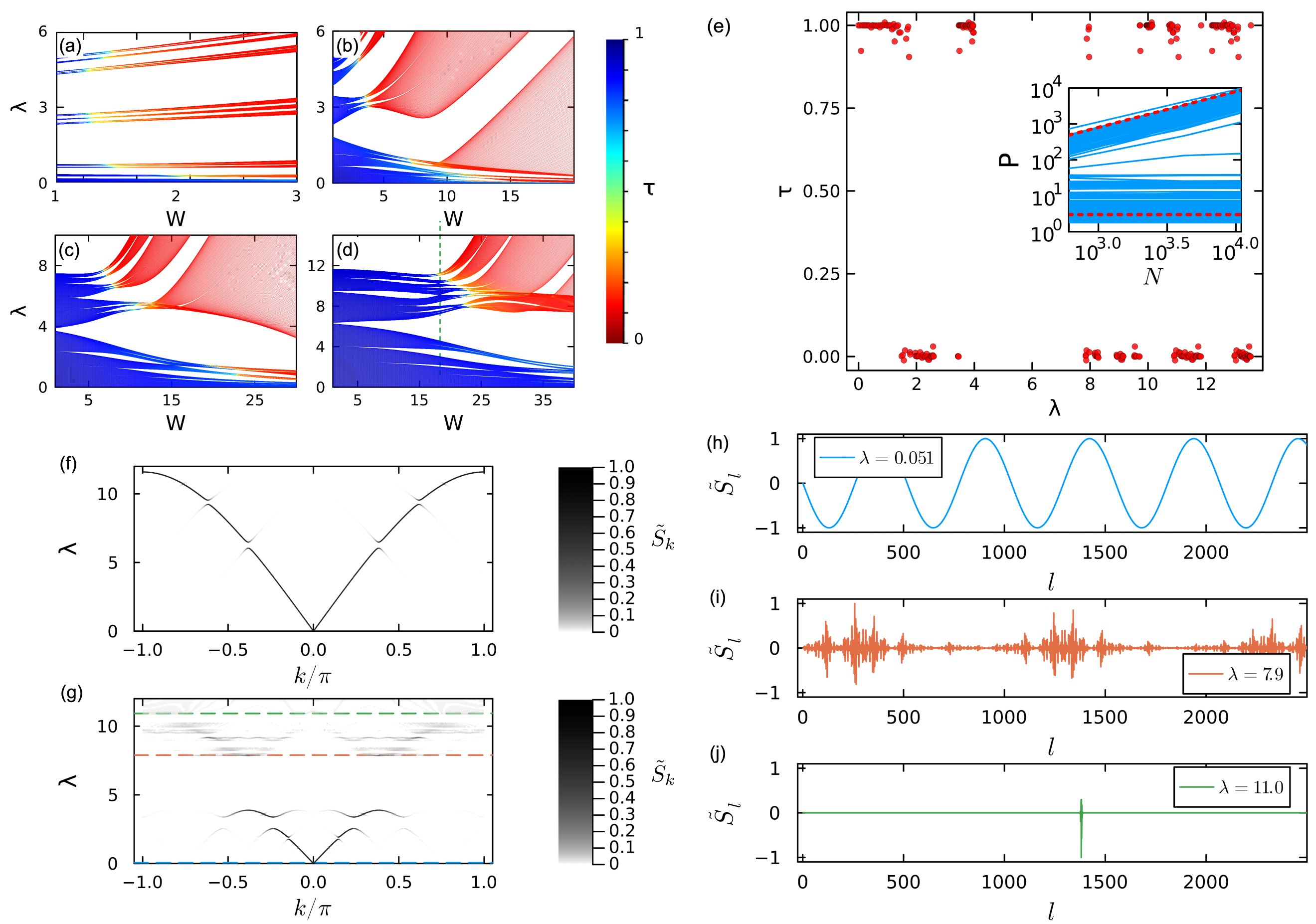}
    \caption{
    Overview of the nonlinear Aubry-André 
    model~\eqref{ham1} properties. In panels (a)-(d), we show elementary excitations spectra for $a = 0.01, 1, 5, 15$ and $N=2584$, respectively, and $\tau$, which indicates the corresponding participation number scaling properties. In panel (e), we present the scaling exponent and participation number $\lambda$-dependencies for $W=15$ [vertical section of panel (d)], which reveals multiple mobility edges. For the scaling analysis we used three system sizes (\(N = 987, 1597, 2584)\), and \(\tau\) averaged over a small energy window \(\delta\lambda = 0.01\). Further insights into the model properties can be obtained using reciprocal space wave functions. In panels (f) and (g), we contrast spectra for $a=3$ with respective $W=3$ and $W=10$. In the former case, all the modes are delocalized, having well-resolved $k$-space peaks, whereas in the latter case, upon $\lambda$ growth, the localization emerges through an intermediate fractal regime. The corresponding wave functions -- extended, critical, and localized -- for $\lambda = 0.051, \, 7.9, \, \text{and} \, 11$ are shown in panels (i)-(k), respectively ($N = 2584$).
    }
    \label{fig:spec}
\end{figure*}

Fluctuations above the ground state, i.e., Bogoliubov -- de Gennes (BdG) modes, are defined using the representation
\be
  \psi_l &=& G_l +  \chi_l e^{-i \lambda t} - \Pi^*_l e^{i \lambda t}. 
\ee
Here $\lambda$ is the energy of an excitation. Inserting $\psi_l$ into Eq.~\eqref{eq_main}, and linearizing the equations of motion in $\chi_l$ and $\Pi_l$ fields, we get the BdG equations:
\be \label{BdG1}
  \lambda \chi_l &=& \phantom{-}J \zeta_l \chi_l + g G^2_l (\chi_l - \Pi_l) - J (\chi_{l+1} +\chi_{l-1}), \\ \label{BdG2}
  \lambda \Pi_l &=& - J \zeta_l \Pi_l + g G^2_l (\chi_l - \Pi_l) + J (\Pi_{l+1} \!+\Pi_{l-1}). \nn  
\ee
Importantly, there is always a solution $\chi_l = \Pi_l \propto G_l $ for $\lambda =0 $ (the Nambu-Goldstone mode due to the global phase invariance breaking). There is also a particle-hole symmetry, so if $\lambda_\nu$ corresponds to an eigenvector $(\chi^\nu_l, \Pi^\nu_l)$ then $-\lambda_\nu$ corresponds to the eigenvector $(\Pi^\nu_l, \chi^\nu_l)$. 

Further simplification can be made by introducing the sum and the difference of the fields $\chi_l = (S_l +D_l)/2$ and  $\Pi_l = (S_l - D_l)/2$. Excluding $D_l$, we have
\be \label{ep1}
  E S_l &\equiv& \frac{\lambda^2}{J} S_l  = \left(J \zeta_l + 2 g G^2_l\right)\left(\zeta_l S_l - S_{l+1} - S_{l-1}\right) \\ &&- J \left( \zeta_{l+1} S_{l+1} + \zeta_{l-1} S_{l-1} -S_{l+2} - 2 S_l - S_{l-2}\right). \nn
\ee
This system of linear equations is suitable for numerical diagonalization for a given ground state (and, consequently, $\zeta_l$). Note that at $W=0$ one has the Bogoliubov spectrum of excitations labeled by momentum $k \in (-\pi,\pi]$,
\be \label{spec1}
  E_k = 4 (1-\cos{k}) \left[ g a  + (1-\cos{k}) J \right],
\ee
and the ``bandwidth'' is $8 (ga + 2 J) $. The latter is mainly determined by $J$ in the low-density regime and $g a$ in the high-density one. Below, we show that, as in the AA model, $W$ should be compared with the bandwidth to judge the localization on a qualitative level.

Localization properties of the eigenstates can be addressed using the participation number, which for the $\nu$-th mode is defined as follows:
\be
  P_\nu = \frac{\left( \sum_l n_{l,\nu} \right)^2}{\sum_l n^2_{l,\nu}}, \quad n_{l,\nu} = |\chi_{l,\nu}|^2 + |\Pi_{l,\nu}|^2.
\ee
The states are classified according to their participation number scaling with the system size. For extended states $P \sim N$, for localized ones $P \sim N^0$. The fractal (critical) modes are characterized by $P \sim N^{\tau}$ with $ 0 < \tau <1 $. Numerically, $\tau$ is determined as the slope of the $\ln{P}(\ln{N})$ function, which is approximately linear for extended and critical states.

Using the equations above, we can develop some intuition on the excitation properties. To do so, we utilize the linear response formula~\eqref{gs} for the ground state. In the low-density regime $ga \ll J$, for high-energy excitations we can neglect $g G^2_l (\chi_l - \Pi_l)$ in Eq.~\eqref{BdG1}. It is easy to see that to the zeroth order in $ga/J$ one returns to the standard AA model, whereas the first correction renormalizes $W$. It reads
\be \label{LowDEst}
  W^\prime = \left[ 1 - \frac{2ga}{J(1-\cos{2 \pi \beta})}\right] W.
\ee
For low-energy excitations, the terms $g G^2_l (\chi_l - \Pi_l)$ in Eqs.~\eqref{BdG1} and~~\eqref{BdG2} are crucial for linear dispersion at $W=0$ and cannot be neglected. So, we can expect that the condition $W^\prime = 2 J$ indicates localization of the highest-energy modes and determines $W_\textrm{NL}$. Using Eq.~\eqref{LowDEst}, one can write
\be \label{LowDEst}
  W_\textrm{NL} \approx 2 J \left[ 1 + \frac{2ga}{J(1-\cos{2 \pi \beta})} \right].
\ee
One of the crucial observations is that this estimation works well only for the states near the edge of the spectrum for $a \ll 1$, see Fig.~\ref{fig:wnl}(a). Counterintuitively, the states in the middle of the spectrum become localized first, which can be understood using the concept of localized phase tongues, see below.

In the high-density regime $g a \gg J$, we can rewrite Eqs.~\eqref{ep1} in the approximate Aubry-André-like form
\be
  \frac{E - 4 g a}{2 g a} S_l &=& \\ &-& S_{l+1} - S_{l-1} - W^\prime(E) \cos{(2 \pi \beta l +\varphi)} S_l \nn
\ee
with the effective AA constant being energy-dependent:
\be \label{MobilityEdgeHighDensity}
  W^\prime (E) =  \left[ \dfrac{E}{2 ga } - 1 +  \cos{(2 \pi \beta)} \right] \frac{W}{2 ga}.
\ee
Using $E_{max} \approx 8 ga$ and condition $ W^\prime (E_{max}) = 2$, we see that localized modes will emerge when 
\be \label{HighDEst}
  W = W_\textrm{NL} \approx \frac{2 g a}{3 + \cos{2\pi \beta}}.
\ee
So, we need $W \sim ga$ to localize high-energy modes and, formally, we are in the regime where the linear response theory for the ground state is inapplicable. However, in Fig.~\ref{fig:wnl}(b) we show that the estimate~\eqref{HighDEst} works on a semi-quantitative level of accuracy. Note also that Eq.~\eqref{MobilityEdgeHighDensity} hints at the existence of the mobility edge in the spectrum for $W > W_{NL}$. Another important feature, the mobility edge $\beta$-dependence, is briefly considered in End Matter.

\begin{figure}
    \centering
    \includegraphics[width=1.0\columnwidth]{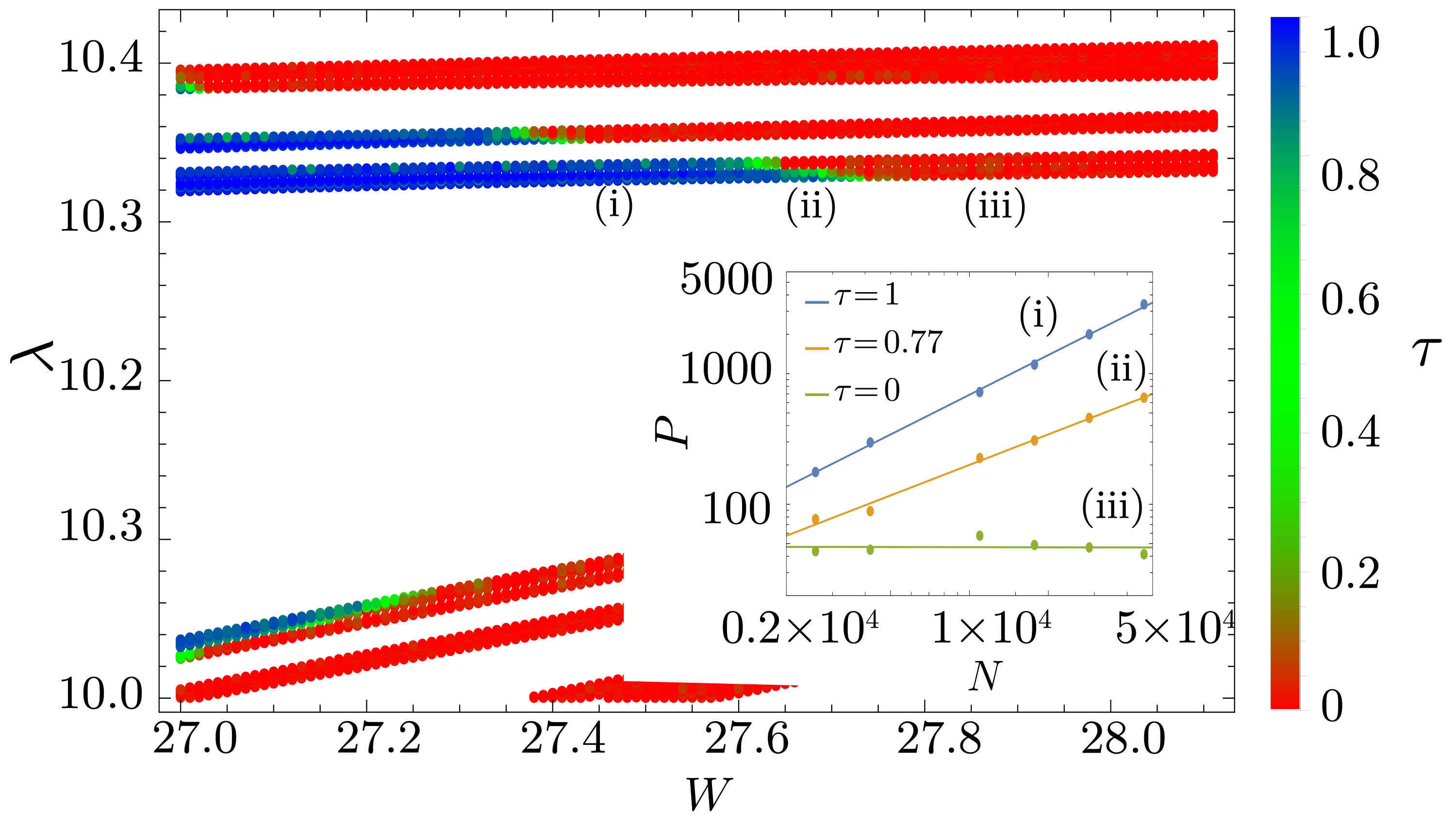}
    \caption{ For a given density $a$ (here, $a=18$) the phase diagram on the $W-\lambda$ plane has nontrivial branching due to the tongues. In the high-density regime, the tongues of the delocalized phase penetrate the localized phase domain. At the tongue ends, fractal states (green dots) can be observed. The inset shows the scaling of the participation ratio for the states indicated by labels (i)-(iii). Exponent $\tau$ reveals their respective extended, fractal, and localized nature. }
    \label{fig:tongues}
\end{figure}

{\it Spectrum and mobility edges}---The complex nature of the BdG spectrum of model~\eqref{ham1} in the case of uncorrelated disorder was shown in Ref.~\cite{kati2021}. The divergence of the localization length was observed not only for low-energy excitations, but also for modes near a certain energy, which depends on the particle density. The nonlinear Aubry-André model also exhibits nontrivial spectral properties, as shown in Fig.~\ref{fig:spec}. In panels (a)-(d), we show the evolution of the BdG excitation spectra with $W$ for various particle densities. The color encodes the fractal dimensions of the respective states. Roughly, one can see that in agreement with the discussion above, the high-density spectra develop the mobility edge at certain values of $W$, which moves to lower energies upon further growth of $W$. Moreover, the lowest-energy modes are delocalized even for the largest $W$ values.

However, a deeper view reveals the fine structure of the spectra for certain intermediate $W$ ranges (depending on $a$). For instance, in Fig.~\ref{fig:spec}(e), we consider the section of the plot~\ref{fig:spec}(d) for $a=15$ along the vertical line $W=23$. While the low- and high-energy modes are delocalized and localized, respectively, for intermediate $\lambda$ we have a mixing of ``subbands'' of localized and delocalized states and, thus, multiple mobility edges. 

For large enough $W$, a single mobility edge separates low-energy (modulated) plane waves and high-energy localized modes [see  Figs.~\ref{fig:spec}(f)-(j)]. Moreover, intermediate critical fractal states can be observed. The difference among the three types of modes is apparent in Figs.~\ref{fig:spec}(h)-(j).

For the high-density regime, multiple mobility edges originate from extended phase tongues, which persist in the form of narrow subbands separating localized phase regions. This idea is illustrated in detail in Fig.~\ref{fig:tongues}.

Physics is also far from being trivial in the low-density regime. Here, the mobility-edge tongues (wedged regions) protrude toward lower \(W\), unlike in the high density regime -- where the tongues protrude toward higher \(W\). They evolve differently at the spectrum’s middle part and edges as  \(a\) increases, as shown in the red (edge) and blue (middle) curves of Fig.~\ref{fig:wnl}(a). The full-spectrum map appears in Fig.~\ref{fig:spec}(a). In particular, states remain extended near the center, localized at intermediate energies, and extended again at the upper edge.

{\it Effective low-energy theory}---Since the pioneering work by Bogoliubov~\cite{bogoliubov1947theory}, it is well-known that in dilute Bose-gas, low-energy excitations essentially differ from the high-energy ones. The former are hydrodynamic phonons (acoustic waves), whereas the latter are nearly free bosons. We observe that in our theory, this property is crucial.   

In our model, the Nambu-Goldstone mode is characterized by $S_l \propto G_l$ and $D_l = 0$. One can guess that smooth low-energy solutions can be obtained by considering variables $\tilde{S}_l = S_l/G_l$ and $\tilde{D}_l = D_l/G_l$. Indeed, we observe that the corresponding excitation real-space profiles look like plain waves even at large $W$ for which the rest of the spectrum is localized, see Fig.~\ref{fig:spec}(h).

To understand the physics behind such a behavior, we rewrite Eq~\eqref{ep1} using $\lambda = i \partial_t$ and gradient expansion of smooth $\tilde{S}_l$. The result can be represented (see End Matter for details) as follows: 
\be
  \partial^2_t \tilde{S} =  (2 J g a + \delta f_l) \partial^2_x \tilde{S} + g_l \partial_x \tilde{S}, 
\ee
with $\delta f_l$ and $g_l$ being zero for $W=0$, so the ``bare'' spectrum is 
\be \label{spec2}
  \lambda^2 = 2 J g a k^2 \equiv c^2 k^2,
\ee
which is just a long-wavelength expansion of Eq.~\eqref{spec1}. For $W \neq 0$, $\delta f_l$ and $g_l$ contain harmonics $\propto W^n \cos{n(2 \pi \beta l +\varphi)}$ and $\sin{n(2 \pi \beta l +\varphi)}$, respectively ($n \geq 1$). However, they are coupled to derivatives of $\tilde{S}$ in the equation of motion. So, after the Fourier transform, we have a  problem similar to the original Aubry-André one:
\be
  (\om^2 - c^2 k^2) \tilde{S}(\om, k) +  \sum_{n \neq 0} \left[ k^2_n f_n + i k_n g_n \right] \tilde{S}(\om, k_n) = 0, \nn \\
\ee
where $k_n = k - 2 \pi n \beta$. The last terms here ``mix'' states with various momenta which differ by $2 \pi \beta$. It is easy to see (using, e.g., the second-order perturbation theory) that the correction means that the effective AA parameter $W^\prime \propto k W$ and no matter how strong the original $W$ is, for small enough energies the BdG modes will be extended.

Numerically, we observe that low-energy excitations are only weakly perturbed, see Figs.~\ref{fig:spec}(f) and (g). However, with the energy growth, the admixture of higher harmonics becomes tangible. Even for strong enough $W$ the whole first ``quasi-band'' (with $ k \in (-\pi \beta, \pi \beta] $ for $\beta < 1/2$  or  $ k \in (-\pi (1-\beta), \pi (1-\beta)] $ for $\beta > 1/2$) can be in the delocalized phase.  This statement is illustrated in Fig.~\ref{fig:spec}(g) for $a=3$ and $W=10$, where states below the largest gap have only several significant Fourier harmonics, and we can ascribe them the momentum in  $k \in (-0.38 \pi, 0.38 \pi)$ interval (here $1-\beta \approx 0.38$).


{\it Conclusion}---We studied the localization properties of the nonlinear Aubry-André model based on the bosonic Gross-Pitaevskii lattice. Unlike the usual Aubry-André model, where there is one critical value $W_\textrm{C} = 2 J$  of the potential strength, which separates regimes of localized and extended states, we observe highly non-trivial behavior of the mobility edge. In the high-density regime in accordance with general arguments, localization first occurs at the edge of the spectrum for high-energy excitations. However, upon further increase of $W$, we reveal peculiar branching of the mobility edge due to the tongues of delocalized phases penetrating the localized phase region. In the low-density regime, counterintuitively, the localization transition first occurs in the middle of the spectrum. This phenomenon is intimately related to the existence of the tongues of the localized phase penetrating the delocalized phase at $W \gtrsim 2$. At the tongues' ends, as a rule, critical (fractal) states can be observed. Finally, the low-energy part of the spectrum is ``protected'' from the localization. This issue can be addressed using the effective low-energy theory, which shows that the corresponding modes feel the renormalized Aubry-André potential.


\textit{Acknowledgements}--- We are grateful to Boris Altshuler, Alexei Andreanov, and Tilen Cadez for valuable discussions. Financial support from the Institute for Basic Science (IBS) in the Republic of Korea through Project No. IBS-R024-D1 is acknowledged.

\bibliography{bib}

\newpage

\begin{center}

\LARGE \bf End Matter
    
\end{center}

\begin{figure}[t]
    \centering
    \includegraphics[width=1.0\linewidth]{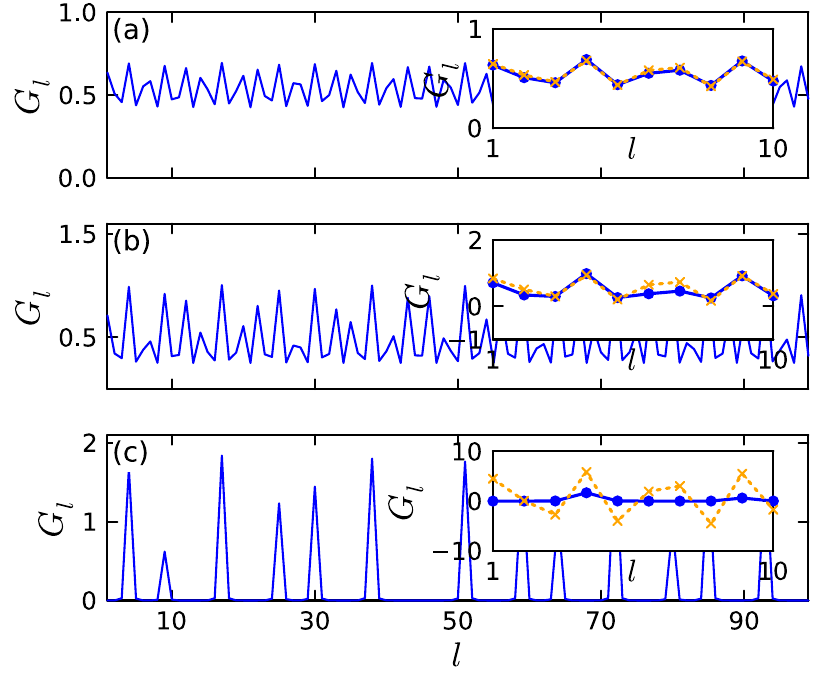}
    \caption{The ground state amplitudes for the density \(a = 0.3\) and $N=2584$ (low-density regime), for different potential strengths \(W = 1, 3, 10\) from (a)-(c). respectively. All notations are identical to those of Fig.~\ref{fig:gs}.}
    \label{fig:gslow}
\end{figure}

\textit{Ground state at low-density regime}---Here we present the results for the ground state, which are complementary to those shown in Fig.~\ref{fig:gs}. In particular, we use $a=0.3$ and three various $W$, see Fig.~\ref{fig:gslow}. The linear response equation~\eqref{gs} works reasonably well for $W \sim 1$ in this case. For the highest $W=10$, the ground state response is essentially nonlinear, which leads to the weakly-coupled condensate puddles.

\textit{Effective low-energy model}---We observed numerically that the eigenvectors of Eq.~\eqref{ep1} recalculated for $\tilde{S}_l = S_l/G_l$ are weakly perturbed plane waves for $E \ll \, \textrm{max}(ga, J)$. So, to obtain an effective theory in the low-energy domain, we can use the gradient expansion. In more detail, we use
\be
  \tilde{S}_{l \pm 1} = \left( \tilde{S} \pm \partial_x \tilde{S} + \frac12 \partial^2_x \tilde{S} \right)\Biggr|_{x=l}.
\ee
After some calculations, Eq.~\eqref{ep1} can be transformed to
\be
  \partial^2_t \tilde{S} = f_l \partial^2_x \tilde{S} + g_l \partial_x \tilde{S}.
\ee
Here 
\be
  f_l &=& J g G^2_l \zeta_l + J^2 \left( 1 + \frac{\zeta^2_l}{2} - \frac{3 \zeta^{(2)}_l}{2}\right), \\
  g_l &=& 2 J g G^2_l \xi_l + J^2 \left( \zeta_l \xi^{(1)}_l - \xi^{(2)}_l \right) \zeta_l
\ee
with
\be 
  \zeta^{(n)}_l = \frac{G_{l+n} + G_{l-n}}{G_l}, \quad \xi^{(n)}_l = \frac{G_{l+n} - G_{l-n}}{G_l}.
\ee
Importantly, at $W = 0 $, $f_l = 2 J g a$ and $g_l =0$, and excitations are acoustic waves with $\lambda^2 = 2 J g a k^2$. More insights can be obtained from the perturbative expansion~\eqref{gs}. For example, in the high-density regime, one has
\be
  f_l &\approx& 2 J g a - (1 + \cos{2 \pi \beta}) \, J W \cos{\left( 2 \pi \beta l + \varphi \right)}, \\
  g_l &\approx& 2 \sin{2 \pi \beta} \, J  W \sin{\left( 2 \pi \beta l + \varphi \right)}.
\ee
 We see, that $f_l$ and $g_l$ are similar to the initial AA potential, which mixes harmonics with $k$ differing by $2 \pi \beta$. In general, perturbations to $f_l$ and $g_l$ contain higher-order harmonics, for example, $\pm 4 \pi \beta$ with coefficient $\propto W^2$ and others.

\begin{figure*}[b]
    \centering
    \includegraphics[width=\linewidth]{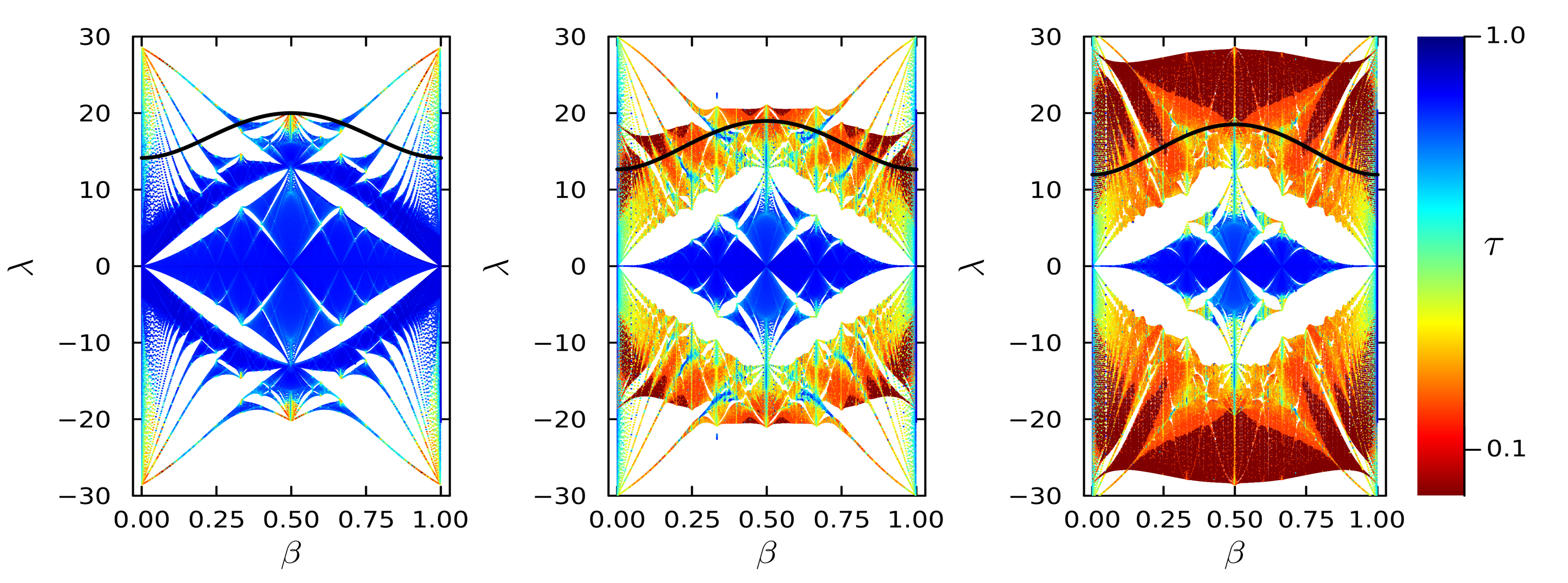}
    \caption{The mobility edge $\beta$-dependence.
    Here, $N = 987$, $a = 50$, $W = 50, \, 62.5, \, 75$ from left to right.
    The black line shows the analytical prediction for the mobility edge at the high-density regime \protect{\eqref{MobilityEdgeHighDensity}}.
    }
    \label{fig:beta_scan}
\end{figure*}

{\it Other values of parameter $\beta$}---We briefly discuss the results for other values of $\beta$. Qualitatively, the above discussion also applies here. Importantly, unlike the AA chain, the mobility edge is expected to depend on $\beta$, as inferred from the low- and high-density approximations~\eqref{LowDEst}, ~\eqref{MobilityEdgeHighDensity}. This allows us to draw colorful Hofstadter butterflies~\cite{hofstadter1976energy} in Fig.~\ref{fig:beta_scan}. It shows the participation number scaling of the BdG spectrum $\lambda$ vs. $\beta$, with the predicted mobility edge \eqref{MobilityEdgeHighDensity} shown by a black line. We find that the mobility edges are indeed $\beta$ dependent. However, as expected, the deviation from the analytical estimation arises because the high-density approximation assumes $g a \gg W$.

\clearpage

\onecolumngrid

\newpage

\begin{center}
    \textbf{\LARGE  Supplementary Material for ``Mobility edges and fractal states in quasiperiodic Gross-Pitaevskii chains'' }
\end{center}

\vskip4mm

\centerline{Oleg I. Utesov, Yeongjun Kim, Sergej Flach} 

\vspace{1cm}

\counterwithin*{figure}{part}
\counterwithin*{equation}{part}
\stepcounter{part}
\renewcommand{\thefigure}{S\arabic{figure}}
\renewcommand{\thetable}{S\arabic{table}}
\renewcommand\theequation{S\arabic{equation}}
\renewcommand\thesection{S\arabic{section}}

\section{Obtaining the ground state}

For the Hamiltonian
\be
    \mathcal{H} = \sum^{N}_{l = 1} \left[ \varepsilon_l |\psi|^2   - J(\psi_l \psi^*_{l+1} + \psi^*_l \psi_{l+1}) + \frac{g}{2}|\psi|^4 \right],
\ee
we need to obtain the normalized ground state with $A$ particles. To do so, we write the normalized wavefunction 
\be
    \tilde{\psi}_l = \frac{1}{\sqrt{A}}\psi_l.
\ee
The Hamiltonian in terms of the normalized wavefunction is
\be
    \tilde{\mathcal{H}} \equiv \mathcal{H}/A = \sum^{N}_{l = 1} \left[ \varepsilon_l |\tilde{\psi}|^2 - J (\tilde{\psi}_l \tilde{\psi}^*_{l+1} + \tilde{\psi}^*_l \tilde{\psi}_{l+1})   +\frac{\tilde{g}}{2} |\tilde{\psi}|^4\right], \nn \\
\ee
where we define $\tilde{g} = gA$.

The ground state has a minimal energy with constraint $\sum_l |\tilde{\psi}_l|^2 = 1$. Corresponding $G_l$'s are obtained when the following gradient becomes zero:
\be
    \nabla \mathcal{L} \equiv &\nabla \left[\mathcal{H} - \mu \left(\sum_l |\tilde{\psi}_l|^2-1\right)\right] = 0.
\ee
This gives the equations
\begin{align}
    (\varepsilon_l - \mu)\psi_l + g|\psi_l|^2\psi_l - J(\psi_{l+1} + \psi_{l-1}) = 0.
\end{align}
The ground state problem can be solved using the gradient descent method, which we implement in the following form:
\be
    (\nabla \mathcal{L})_t = \frac{\nabla \mathcal{L} -(\nabla \mathcal{L} \cdot \psi)\psi}{\left\|\nabla \mathcal{L} -(\nabla \mathcal{L} \cdot \psi)\psi \right\|}.
\ee
Here, \((*)_t\) and \((*)_n\) are parallel, and normal components of the gradient to the unit sphere of norm 1. The gradient descent must be restricted to this unit sphere. The update is done (while conserving its norm) as
\be
    x^{(k+1)} = \cos (\phi_k) x^{(k)} - \sin (\phi_k) \nabla \mathcal{L}.
\ee
with a parameter step length $\phi_k$ as in the conventional gradient descent method. At the ground state, the tangent component $(\nabla \mathcal{H})_t$ becomes zero; on the other hand, the normal component satisfies
\be
    (\nabla \mathcal{H})_n = \mu\nabla \| \tilde{\psi} \|^2 = 2\mu \vec{\psi}.
\ee



\section{Extended Harper model}

This model was studied, e.g., in Refs.~\cite{ino2006critical,gong2008fidelity,avila2017spectral}. Here we will mostly follow the notations of Ref.~\cite{avila2017spectral}.

We show that for the high density regime, the equation describing BdG excitations (Eq.~11 of the main text) corresponds to the extended Harper model (EHM) one. Importantly, this property hints at the possibility of observing fractal states in our model. 

At high density, we can write,
\be \label{BdG_At_HighDensity}
    E S_l = 2gG^2_l(\zeta_l S_l - S_{l+1} - S_{l-1})
\ee
Let us define the following operators
\be
    \notag \hat{D} &=& \sum_l 2gG^2_l \ketbra{l}{l}, \\
    \hat{H}_0 &=& \sum_l \zeta_l\ketbra{l}{l} -\ketbra{l}{l+1} -\ketbra{l+1}{l}.
\ee
Then, we can write Eq.~\eqref{BdG_At_HighDensity} as follows
\begin{align}
    E\ket{S} = (\hat{D}\hat{H}_0)\ket{S}.
\end{align}
where $\ket{S} = \sum_l S_l\ket{l}$.
Performing the similarty transformation with $\hat{D}^{1/2}$ leads to effective BdG Hamiltonian $\hat{D}^{1/2}\hat{H}_0\hat{D}^{1/2}$, which is Hermitian.
So, after this transformation, the BdG equations are given by
\begin{align} \label{Aspec}
    E \tilde{S}_l = (h_l + h_{l-1})\tilde{S}_l - h_l\tilde{S}_{l+1} - h_{l-1}\tilde{S}_{l-1}.
\end{align}
where $\tilde{S}_l = (2g)^{-1/2}S_l/G_l$, and
$h_l = 2gG_lG_{l+1}$, which approximately reads
\begin{align}
    h_l = 2ga - W \cos(\pi\beta)(e^{2\pi i(\beta l + \beta/2)} + e^{-2\pi i(\beta l + \beta/2)}).
\end{align}
With an appropriate rescaling, we have the extended Harper model with parameters given by:
\begin{gather}
    \notag \lambda_2 = \frac{ga}{W\cos^2(\pi\beta)}, \\
    \lambda_1=\lambda_3 = \frac{1}{2|\cos(\pi\beta)|}.
\end{gather}
The phase diagram of the extended Harper model is known exactly~\cite{avila2017spectral}.

In our model, we obtain
\begin{align}
    \notag ga &> W|\cos(\pi\beta)| \quad \mathrm{extended}\\
    ga  &< W|\cos(\pi\beta)| \quad \mathrm{critical}.
\end{align}
These inequalities suggest the existence of fractal states in the spectrum of the nonlinear Aubry-Andre model; however, in the range of $W$ where the whole procedure leading to Eq.~\eqref{Aspec} is questionable (it relies on the $W \ll ga$ condition). Moreover, from the inequalities above, we cannot obtain any information related to mobility edges and localized states.

\begin{figure*}[t]
    \centering
    \includegraphics[width=0.45\linewidth]{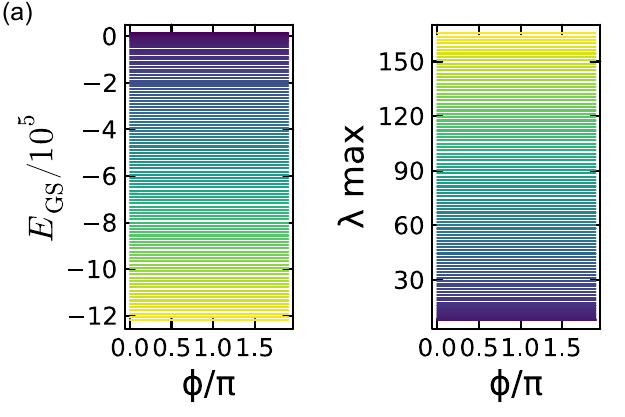}
     \includegraphics[width=0.45\linewidth]{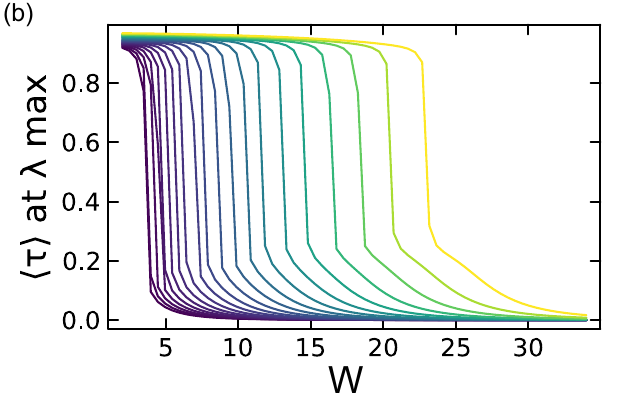}
    \includegraphics[width=0.45\linewidth]{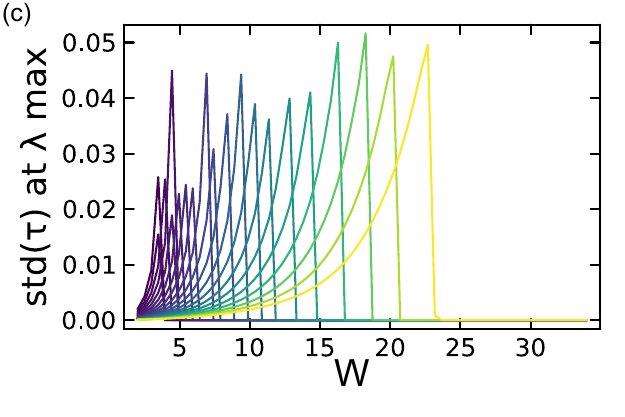}
    \includegraphics[width=0.45\linewidth]{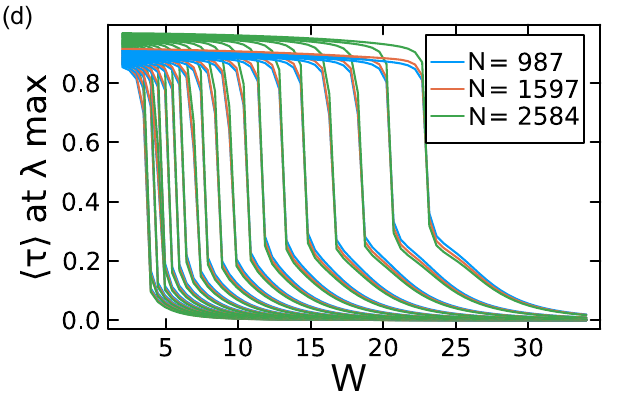}
    \caption{Here the range of $a \in (2,20)$ is scanned (high-density regime), which corresponds to colors from violet to yellow in panels (a)-(c). (a) There is almost no $\phi$-induced variation for the ground state energy and maximal eigenvalue. (b) Averaged over $\phi$ participation ratio exponent $\tau$ clearly indicates localization transition at the edge of the spectrum. (c) Standard deviation of $\tau$ essentially increases at near the transition point. (d) Scaling of the participation ratio exponent with the system size; for curve bunches from left to right, $a$ changes from 2 to 20.}
    \label{fig:supp1}
\end{figure*}

\section{Finite-size scaling, rational approximations, and phase}
 
In our calculations throughout the manuscript, we use
chains of finite size with periodic boundary conditions for the Aubry-Andre potential
\be
  \varepsilon_l = W \cos{(2 \pi \beta l +\varphi)}
\ee
Since we are mostly interested in $\beta = (\sqrt{5}-1)/2$, it is natural to use its rational approximation in the form of ratios of Fibonacci numbers
\be
  \beta \approx \frac{F_{n-1}}{F_n},
\ee
which suggests using, e.g.,
\be
  \beta = \frac{610}{987}, \,  \frac{1597}{2584}, \, \frac{6765}{10946}, ... 
\ee
The denominators here represent natural choices for the system sizes: $N=987, 2584, ...$ This set of system sizes allows us to check scaling properties of eigenstates and to determine corresponding fractal dimensions (see Fig.~\ref{fig:supp1}).

Another important issue for the Aubry-Andre problem is the possible dependence of physics on the phase variable $\varphi$ (see, e.g., Ref.~\cite{morales2014}). In Fig.~\ref{fig:supp1}, we show this is not the case for our model. Even though near the transition point the standard deviation of participation ratio exponent $\tau$ slightly increases,  we clearly observe the change of behavior from extended to localized for high-energy modes at $W_\textrm{NL}$.

\end{document}